\documentclass[aps,prd,twocolumn,superscriptaddress,showpacs]{revtex4}
\newlength{\plotwidth}
\usepackage{graphicx}
\usepackage{subfigure}
\usepackage{amsmath}
\usepackage{amssymb}
\usepackage{dcolumn}
\usepackage{booktabs}
\usepackage{psfrag}
\usepackage{xspace}
\usepackage{html}
\usepackage{setspace}
\usepackage{color}

\newcommand\etal{\it et al.}
\newcommand{\qbar}{\ensuremath{\bar{q}}\xspace}

\newcommand{\tbar}{\ensuremath{\bar{t}\xspace}}

\newcommand{\pbar}{\ensuremath{\bar{p}\xspace}}

\newcommand{\ttbar}{\ensuremath{{t\tbar}}\xspace}
\newcommand{\qqbar}{\ensuremath{{q\qbar}}\xspace}
\newcommand{\ppbar}{\ensuremath{{p\pbar}}\xspace}
\newcommand{\ee}{\ensuremath{{e^{+}e^{-}}}\xspace}
\newcommand{\mm}{\ensuremath{{\mu^{+}\mu^{-}}}\xspace}

\newcommand{\tZq}{\ensuremath{{t\rightarrow Zq}}\xspace}
\newcommand{\tZc}{\ensuremath{{t\rightarrow Zc}}\xspace}
\newcommand{\tZu}{\ensuremath{{t\rightarrow Zu}}\xspace}
\newcommand{\tWb}{\ensuremath{{t\rightarrow Wb}}\xspace}
\newcommand{\zsig}{\ensuremath{{Z_\textrm{signal}}}\xspace}
\newcommand{\ztag}{\ensuremath{{Z_\textrm{tagged}}}\xspace}
\newcommand{\zanti}{\ensuremath{{Z_\textrm{non-tagged}}}\xspace}
\newcommand{\zcont}{\ensuremath{{Z_\textrm{control}}}\xspace}
\newcommand{\ftag}{\ensuremath{{f_\textrm{tag}}}\xspace}

\newcommand{\sigjes}{\ensuremath{\sigma_\textrm{JES}}}
\newcommand{\btZq}{\ensuremath{{\mathcal B}(\tZq)}\xspace}
\newcommand{\btZc}{\ensuremath{{\mathcal B}(\tZc)}\xspace}
\newcommand{\btZu}{\ensuremath{{\mathcal B}(\tZu)}\xspace}
\newcommand{\btWb}{\ensuremath{{\mathcal B}(\tWb)}\xspace}
\newcommand{\Rsig}{\ensuremath{{\mathcal{R}}_{\mathrm{sig}}}\xspace}

\newcommand{\pt}{\ensuremath{p_{T}}\xspace}

\newcommand{\et}{\ensuremath{E_{T}}\xspace}
\newcommand{\Et}{\et}

\newcommand{\mt}{\ensuremath{m_{T}}\xspace}

\newcommand{\chisq}{\ensuremath{\chi^2}\xspace}
\newcommand{\sqchisq}{\ensuremath{\sqrt{\chi^2}}\xspace}

\newcommand{\unit}[1]{\ensuremath{\;\mathrm{#1}}\xspace}

\newcommand{\gev}{\unit{GeV}}
\newcommand{\gevc}{\ensuremath{\unit{GeV}/c}\xspace}
\newcommand{\gevcsq}{\ensuremath{\unit{GeV}/c^{2}}\xspace}
\newcommand{\tev}{\unit{TeV}}
\newcommand{\invfb}{\unit{fb^{-1}}}

\newcommand{\pyth}{{\sc pythia}\xspace}
\newcommand{\alp}{{\sc alpgen}\xspace}

\newcommand{\Zj}{\ensuremath{Z}+jets\xspace}

\begin{document}
\preprint{PRL Draft Version 2.0, CDF/PHYS/TOP/CDFR/9266}

\title{\bf Search for the Flavor Changing Neutral Current Decay 
  \boldmath\tZq\\in \ppbar Collisions at $\sqrt{s}=1.96$\tev\unboldmath}

\date{\today}
\affiliation{Institute of Physics, Academia Sinica, Taipei, Taiwan 11529, Republic of China} 
\affiliation{Argonne National Laboratory, Argonne, Illinois 60439} 
\affiliation{University of Athens, 157 71 Athens, Greece} 
\affiliation{Institut de Fisica d'Altes Energies, Universitat Autonoma de Barcelona, E-08193, Bellaterra (Barcelona), Spain} 
\affiliation{Baylor University, Waco, Texas  76798} 
\affiliation{Istituto Nazionale di Fisica Nucleare Bologna, $^t$University of Bologna, I-40127 Bologna, Italy} 
\affiliation{Brandeis University, Waltham, Massachusetts 02254} 
\affiliation{University of California, Davis, Davis, California  95616} 
\affiliation{University of California, Los Angeles, Los Angeles, California  90024} 
\affiliation{University of California, San Diego, La Jolla, California  92093} 
\affiliation{University of California, Santa Barbara, Santa Barbara, California 93106} 
\affiliation{Instituto de Fisica de Cantabria, CSIC-University of Cantabria, 39005 Santander, Spain} 
\affiliation{Carnegie Mellon University, Pittsburgh, PA  15213} 
\affiliation{Enrico Fermi Institute, University of Chicago, Chicago, Illinois 60637} 
\affiliation{Comenius University, 842 48 Bratislava, Slovakia; Institute of Experimental Physics, 040 01 Kosice, Slovakia} 
\affiliation{Joint Institute for Nuclear Research, RU-141980 Dubna, Russia} 
\affiliation{Duke University, Durham, North Carolina  27708} 
\affiliation{Fermi National Accelerator Laboratory, Batavia, Illinois 60510} 
\affiliation{University of Florida, Gainesville, Florida  32611} 
\affiliation{Laboratori Nazionali di Frascati, Istituto Nazionale di Fisica Nucleare, I-00044 Frascati, Italy} 
\affiliation{University of Geneva, CH-1211 Geneva 4, Switzerland} 
\affiliation{Glasgow University, Glasgow G12 8QQ, United Kingdom} 
\affiliation{Harvard University, Cambridge, Massachusetts 02138} 
\affiliation{Division of High Energy Physics, Department of Physics, University of Helsinki and Helsinki Institute of Physics, FIN-00014, Helsinki, Finland} 
\affiliation{University of Illinois, Urbana, Illinois 61801} 
\affiliation{The Johns Hopkins University, Baltimore, Maryland 21218} 
\affiliation{Institut f\"{u}r Experimentelle Kernphysik, Universit\"{a}t Karlsruhe, 76128 Karlsruhe, Germany} 
\affiliation{Center for High Energy Physics: Kyungpook National University, Daegu 702-701, Korea; Seoul National University, Seoul 151-742, Korea; Sungkyunkwan University, Suwon 440-746, Korea; Korea Institute of Science and Technology Information, Daejeon, 305-806, Korea; Chonnam National University, Gwangju, 500-757, Korea} 
\affiliation{Ernest Orlando Lawrence Berkeley National Laboratory, Berkeley, California 94720} 
\affiliation{University of Liverpool, Liverpool L69 7ZE, United Kingdom} 
\affiliation{University College London, London WC1E 6BT, United Kingdom} 
\affiliation{Centro de Investigaciones Energeticas Medioambientales y Tecnologicas, E-28040 Madrid, Spain} 
\affiliation{Massachusetts Institute of Technology, Cambridge, Massachusetts  02139} 
\affiliation{Institute of Particle Physics: McGill University, Montr\'{e}al, Canada H3A~2T8; and University of Toronto, Toronto, Canada M5S~1A7} 
\affiliation{University of Michigan, Ann Arbor, Michigan 48109} 
\affiliation{Michigan State University, East Lansing, Michigan  48824}
\affiliation{Institution for Theoretical and Experimental Physics, ITEP, Moscow 117259, Russia} 
\affiliation{University of New Mexico, Albuquerque, New Mexico 87131} 
\affiliation{Northwestern University, Evanston, Illinois  60208} 
\affiliation{The Ohio State University, Columbus, Ohio  43210} 
\affiliation{Okayama University, Okayama 700-8530, Japan} 
\affiliation{Osaka City University, Osaka 588, Japan} 
\affiliation{University of Oxford, Oxford OX1 3RH, United Kingdom} 
\affiliation{Istituto Nazionale di Fisica Nucleare, Sezione di Padova-Trento, $^u$University of Padova, I-35131 Padova, Italy} 
\affiliation{LPNHE, Universite Pierre et Marie Curie/IN2P3-CNRS, UMR7585, Paris, F-75252 France} 
\affiliation{University of Pennsylvania, Philadelphia, Pennsylvania 19104} 
\affiliation{Istituto Nazionale di Fisica Nucleare Pisa, $^q$University of Pisa, $^r$University of Siena and $^s$Scuola Normale Superiore, I-56127 Pisa, Italy} 
\affiliation{University of Pittsburgh, Pittsburgh, Pennsylvania 15260} 
\affiliation{Purdue University, West Lafayette, Indiana 47907} 
\affiliation{University of Rochester, Rochester, New York 14627} 
\affiliation{The Rockefeller University, New York, New York 10021} 

\affiliation{Istituto Nazionale di Fisica Nucleare, Sezione di Roma 1, $^v$Sapienza Universit\`{a} di Roma, I-00185 Roma, Italy} 

\affiliation{Rutgers University, Piscataway, New Jersey 08855} 
\affiliation{Texas A\&M University, College Station, Texas 77843} 
\affiliation{Istituto Nazionale di Fisica Nucleare Trieste/\ Udine, $^w$University of Trieste/\ Udine, Italy} 
\affiliation{University of Tsukuba, Tsukuba, Ibaraki 305, Japan} 
\affiliation{Tufts University, Medford, Massachusetts 02155} 
\affiliation{Waseda University, Tokyo 169, Japan} 
\affiliation{Wayne State University, Detroit, Michigan  48201} 
\affiliation{University of Wisconsin, Madison, Wisconsin 53706} 
\affiliation{Yale University, New Haven, Connecticut 06520} 
\author{T.~Aaltonen}
\affiliation{Division of High Energy Physics, Department of Physics, University of Helsinki and Helsinki Institute of Physics, FIN-00014, Helsinki, Finland}
\author{J.~Adelman}
\affiliation{Enrico Fermi Institute, University of Chicago, Chicago, Illinois 60637}
\author{T.~Akimoto}
\affiliation{University of Tsukuba, Tsukuba, Ibaraki 305, Japan}
\author{M.G.~Albrow}
\affiliation{Fermi National Accelerator Laboratory, Batavia, Illinois 60510}
\author{B.~\'{A}lvarez~Gonz\'{a}lez}
\affiliation{Instituto de Fisica de Cantabria, CSIC-University of Cantabria, 39005 Santander, Spain}
\author{S.~Amerio$^u$}
\affiliation{Istituto Nazionale di Fisica Nucleare, Sezione di Padova-Trento, $^u$University of Padova, I-35131 Padova, Italy} 

\author{D.~Amidei}
\affiliation{University of Michigan, Ann Arbor, Michigan 48109}
\author{A.~Anastassov}
\affiliation{Northwestern University, Evanston, Illinois  60208}
\author{A.~Annovi}
\affiliation{Laboratori Nazionali di Frascati, Istituto Nazionale di Fisica Nucleare, I-00044 Frascati, Italy}
\author{J.~Antos}
\affiliation{Comenius University, 842 48 Bratislava, Slovakia; Institute of Experimental Physics, 040 01 Kosice, Slovakia}
% ...

\collaboration{CDF Collaboration\footnote{With visitors from $^a$Universiteit Antwerpen, B-2610 Antwerp, Belgium, 
$^b$Chinese Academy of Sciences, Beijing 100864, China, 
$^c$University of Bristol, Bristol BS8 1TL, United Kingdom, 
$^d$University of California Irvine, Irvine, CA  92697, 
$^e$University of California Santa Cruz, Santa Cruz, CA  95064, 
$^f$Cornell University, Ithaca, NY  14853, 
$^g$University of Cyprus, Nicosia CY-1678, Cyprus, 
$^h$University College Dublin, Dublin 4, Ireland, 
$^i$University of Edinburgh, Edinburgh EH9 3JZ, United Kingdom, 
$^j$Universidad Iberoamericana, Mexico D.F., Mexico, 
$^k$University of Manchester, Manchester M13 9PL, England, 
$^l$Nagasaki Institute of Applied Science, Nagasaki, Japan, 
$^m$University de Oviedo, E-33007 Oviedo, Spain, 
$^n$Queen Mary, University of London, London, E1 4NS, England, 
$^o$Texas Tech University, Lubbock, TX  79409, 
$^p$IFIC(CSIC-Universitat de Valencia), 46071 Valencia, Spain,
$^x$Royal Society of Edinburgh/Scottish Executive Support Research Fellow, 
}}
\noaffiliation

\author{The CDF Collaboration}

\begin{abstract}

  We report a search for the flavor changing neutral current (FCNC)
  decay of the top quark \tZq ($q=u,c$) in \ppbar collisions at
  $\sqrt{s} = 1.96\tev$ using a data sample corresponding to an
  integrated luminosity of 1.9\invfb collected by the CDF II
  detector. This decay is strongly suppressed in the standard model
  (SM) and an observation of a signal at the Tevatron would be an
  indication of physics beyond the SM. Using $Z+\ge4$ jet final state
  candidate events, both with and without an identified bottom quark
  jet, we discriminate signal from background by exploiting kinematic
  constraints present in FCNC events and obtain an upper limit of
  $\btZq < 3.7\%$ at 95\% C.L.

\end{abstract}



\maketitle

%
%
~\\ \hspace*{8pt} 
Flavor changing neutral current (FCNC) interactions, which mediate
transitions from one type of quark to another with the same electric
charge, are suppressed in the standard model of particle physics
(SM). FCNC processes are therefore sensitive indicators of physics
beyond the SM (BSM). Presently there are only loose experimental
bounds on FCNC decays of the $t$ (top) quark~\cite{Fritzsch:1989qd},
the heaviest known quark.  While the SM branching fraction for \tZq
($q=u,c$) is predicted to be
$\mathcal{O}(10^{-14})$~\cite{Aguilar-Saavedra:2004wm}, BSM models
such as supersymmetry and quark compositeness allow branching
fractions as high as
$\mathcal{O}(10^{-4})$~\cite{Aguilar-Saavedra:2004wm, Larios:2006pb}.
An observation of the top FCNC decay \tZq with present-size data
samples would be a strong indication of BSM physics.

In Tevatron Run~I, a CDF search for the \tZq decay yielded the
branching fraction upper limit $\btZq < 33\%$ (95\%
C.L.)~\cite{RunIfcnc}. The current best 95\% C.L.\ upper limit on
$\btZq$, 13.7\%, was set by the L3 experiment~\cite{L3fcnc} from the
non-observation of FCNC single $t$ quark production.  These analyses
assumed a $t$ quark mass of $m_t=175\gevcsq$, close to the current
world average of $m_t=172.4\pm 1.2 \gevcsq$~\cite{TopMass2008}. In
this Letter we also assume $m_t=175\gevcsq$ and evaluate the small
effect of different $m_t$ on the result.

We present the first Tevatron Run~II search for the top FCNC decay
\tZq in \ttbar pairs. The data sample analyzed for this Letter
corresponds to an integrated luminosity of 1.9\invfb
of~\ppbar~collisions at $\sqrt{s}=1.96\tev$, collected by the CDF~II
detector from March~2002 to May~2007. Our primary signal signature
consists of the FCNC decay \tZq ($q=u,c$) together with the dominant
SM decay $\bar t\to W^- \bar b$ (charge conjugate modes are
implied). We search for the $Z$~boson via its decays $Z\to\ee$ and
$Z\to\mm$ and for the decay of the $W$~boson into a quark-antiquark
pair. Two leptons and at least four jets of hadrons, coming from the
four quarks from secondary decays, can be observed in the detector. We
also allow for events in which both $t$ quarks undergo FCNC decays.
We consider several SM background processes that also result in final
states with a reconstructed $Z$ boson and four or more jets. The
dominant background process is the production of $Z$ bosons with
associated jets (\Zj). Smaller contributions come from SM \ttbar
production and the production of pairs of gauge bosons (dibosons),
$WZ$ and $ZZ$. The contributions from $WW$ diboson production and from
$W$ bosons produced in association with jets are negligible.

%
%
The components of the CDF II detector relevant to this analysis are
briefly described here; a more complete description can be found
elsewhere~\cite{CDF}.  The transverse momenta \pt and pseudorapidities
$\eta$~\cite{cdfdefs} of charged particles are measured by a silicon
strip detector~\cite{SIstrip} and a 96-layer drift chamber
(COT)~\cite{COT} inside a $1.4\unit{T}$ solenoidal magnetic field.
The silicon detector and the COT provide good combined reconstruction
efficiency for $|\eta|<1.2$. The precise position resolution of the
silicon detector is also crucial for identifying displaced secondary
vertices from from long-lived $B$ hadrons ($b$-tagging).
Electromagnetic (EM) and hadronic calorimeters measure energies of
charged and neutral particles in the central ($|\eta|<1.1$) and
end-plug ($1.1<|\eta|<3.6$) regions.  Drift chambers and scintillation
counters located on the rear of the calorimeters and behind an
additional steel absorber detect muons with $|\eta|<1.0$.  A three
level trigger selects events that contain electrons (muons) with
$\et>18\gev$ ($\pt>18\gevc$).

%
%
We use the \pyth v6.216 Monte Carlo (MC) generator~\cite{Pythia} to
simulate the FCNC signal in \ttbar events and all sources of SM
background, except \Zj production, whose kinematic distributions we
simulate with the \alp MC generator~\cite{Alpgen}, v2.10$^\prime$
interfaced to \pyth v6.325.  Acceptance, efficiency, and kinematic
distributions of signal and background are determined from the above
MC simulations. We add a \tZq decay channel to \pyth by forcing the
$Z$ helicity to be consistent with the expectation from an SM-like
Higgs mechanism of 65\% longitudinally polarized and 35\% left-handed
$Z$ bosons.  The main signal MC sample contains events that decay
according to our primary signature, $\ttbar \to Zq\,Wb$ with $q=c$. If
the \tZc decay is replaced by \tZu, the probability to $b$-tag an FCNC
signal event is reduced from 50\% to 45\%, where we apply the SecVtx
algorithm~\cite{CDF} for the $b$-tagging. We have also generated an MC
event sample in which both $t$ quarks decay via the FCNC mode and the
$Z$ is allowed to decay into the \ee, \mm, and \qqbar decay modes.
The amount of $\ttbar\to Zc\,Zc$ relative to $\ttbar\to Wb\,Zc$ is
adjusted according to the branching fraction \btZq during the limit
calculation.

We apply corrections for trigger efficiencies and for different
efficiencies in data and simulation for lepton identification and
reconstruction, $b$-tagging, and misidentification of secondary
vertices (mistags).

\begin{figure}[t!]
\vspace*{-0.2in}
\centerline{\includegraphics[width=\plotwidth]{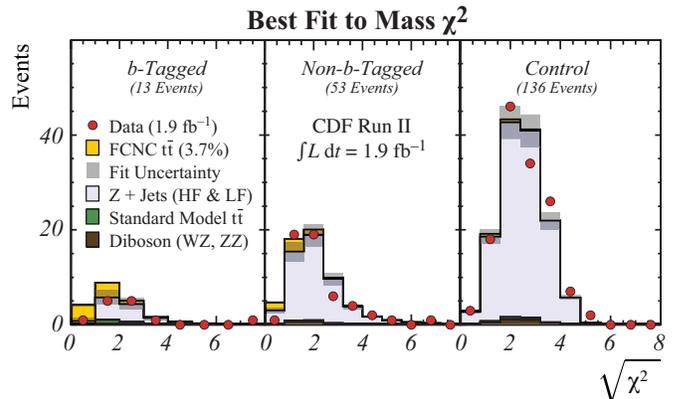} }
\vspace*{-0.2in}
\caption{\label{fig:moneyplot} Mass \chisq distribution for $b$-tagged
  and non-$b$-tagged signal regions and the control region. The data
  points as a function of \sqchisq are compared to the
  SM background prediction and the expected FCNC yield at the observed
  95\% C.L.\ upper limit on the branching fraction $\btZq <
  3.7\%$. The data are consistent with the background prediction.}
\vspace*{-0.2in}
\end{figure}

%
The analysis proceeds in three steps. Our base selection requires a
$Z$ boson and four or more jets. An optimized event selection
separates the FCNC signal from the SM backgrounds based on kinematic
properties of the event. All events that pass the optimized selection
are further divided into two signal regions based on whether any of
the four leading jets has an identified $b$-jet (``$b$-tagged
region'') or not (``non-$b$-tagged region''); those events that pass
the base selection but not the optimized selection are placed in a
``control'' region. While only 12\% of FCNC signal acceptance falls in
the control region, two-thirds of the SM backgrounds are located
there. We finally fit the data to one-dimensional probability density
histograms (``templates'') of one of the kinematic distributions (mass
\chisq, see below) to derive a limit on \btZq. This is illustrated in
Fig.~\ref{fig:moneyplot}. To reduce systematic uncertainties, the
measurement is normalized to the measured event yield in SM \ttbar
production.

%
%
 Our base selection chooses events with two oppositely-charged
leptons of the same flavor ($e$ or $\mu$) and four or more jets. One
lepton is required to be a central electron or muon, the other can be
a forward electron or a track, as described below. The leptons must be
compatible with originating from a $Z$ boson in the mass window from
76\gevcsq to 106\gevcsq ($>3\sigma$ in units of CDF's $Z$ mass
resolution).

All leptons used in this analysis are required to be well isolated in
a cone of $\Delta R =\sqrt{(\Delta\eta)^2 + (\Delta\phi)^2} <
0.4$~\cite{cdfdefs} and to have transverse energies \Et (momenta \pt)
greater than 20\gev (20\gevc).  Electrons are identified by requiring
an energy cluster in the EM calorimeter with a single track pointing
to it. Central electrons are required to have a high quality COT-based
track, calorimeter cluster \Et consistent with the track \pt, a high
fraction of the total energy deposition in the EM calorimeter, and a
lateral shower profile consistent with electron showers. Forward
electrons are reconstructed in the end-plug calorimeter and have
similar constraints except that the tracks are reconstructed only in
the silicon detector and the calorimeter cluster \Et and track \pt are
not compared. Muons are identified by matching tracks reconstructed in
the COT to track segments reconstructed in the muon chambers and by
requiring energy depositions in the calorimeters consistent with
minimum ionizing particles. The muons in this analysis are required to
be in the central region ($|\eta| < 1.0$). We double the acceptance
for leptonic $Z$ decays by allowing one of the two lepton candidates
to satisfy weaker selection criteria, requiring only an isolated track
which passes COT and silicon detector track quality cuts. For tracks
used as electrons, if an EM calorimeter tower is associated to the
track and the energy of the EM tower is greater than the track
momentum, the EM tower energy is used instead of the track momentum.

Jets are identified by energy deposited in the calorimeters within a
cone of $\Delta R < 0.4$.  To improve the parton energy estimate, jets
are corrected for instrumental effects~\cite{JES}.  We select events
with at least four jets with corrected $\Et > 15\gev$ and $|\eta| <
2.4$. 

%
%
We separate the FCNC signal from the MC background with the help of
further selection criteria in addition to the above base selection
that will form the optimized selection, a mass \chisq, the transverse
mass of the system, and the \Et of the four leading (highest \Et)
jets.

The decay $\ttbar\to Wb\,Zq \to q \bar{q}'b\,\ell\ell q''$ contains no
high-energy neutrinos; therefore we can fully reconstruct the event
kinematics.  The four jets in signal events result from the $b$ quark
and the decay products of the $W$ in the \tWb decay, and the $c$ or
$u$ quark from the \tZq decay. We form all permutations of the four
leading jets in the events to compare the reconstructed masses
($m_\textrm{rec}$) of the $W$, top quark decaying to $Wb$, and top
quark decaying to $Zq$. We define a mass \chisq as
\begin{eqnarray}
\chi^2 = \left(\frac{m_{W,\mathrm{rec}} - m_{W}}{\sigma_{W}} \right)^2 
&+& \left(\frac{m_{\tWb,\mathrm{rec}} - m_{t}}{\sigma_{\tWb}} \right)^2 \nonumber \\
&+& \left(\frac{m_{\tZq,\mathrm{rec}} - m_{t}}{\sigma_{\tZq}} \right)^2,
\label{eq:chisquare}
\end{eqnarray}
and select the permutation with the smallest \chisq. We scale the
measured four-momenta of the $W$ and $Z$ boson daughter particles such
that the boson masses are fixed to the world average values~\cite{PDG}
and use the scaled four-momenta to calculate the two top quark
masses. The widths used are given by the standard deviations of the
reconstructed masses measured in the MC simulation of FCNC
events. Using the correct pairing of jets to partons, we extract
$\sigma_{W}=15\gevcsq$, $\sigma_{\tWb}=24\gevcsq$, and
$\sigma_{\tZq}=21\gevcsq$. We expect FCNC signal events to populate
the low \chisq region and background events to result in higher \chisq,
see Fig.~\ref{fig:moneyplot}.  We have verified that the components of
Eq.~(\ref{eq:chisquare}) describe the data well in events with a $Z$
boson and three jets. 

Since the FCNC signal events originate from \ttbar decays, they
contain more central $Z$ bosons and jets than background events. To
exploit this, we use the transverse mass of the $Z$ and the four
leading jets, defined as $\mt = \sqrt{\left(\sum\et \right)^2 -
  \left(\sum\vec{p}_{T}\right)^2 }$, as a selection criterion. We also
apply a tiered cut on the \Et of the four leading jets, as FCNC signal
events contain jets with higher transverse momenta than SM background
events.

We optimized these additional selection criteria for the best expected
limit on \btZq in the absence of a signal, using the MC simulation and
a signal-depleted control region in the data ($\sqrt{\chisq}>3$). The
optimization was performed in the blind phase of a counting experiment
analysis using the first 1.1\invfb of integrated
luminosity~\cite{V1result}. We leave the optimized selection unchanged
for the full 1.9\invfb result; it requires transverse mass $\mt\ge
200\gevcsq$, leading jet $\Et\ge 40\gev$, second jet $\Et\ge 30\gev$,
third jet $\Et\ge 20\gev$, and fourth jet $\Et\ge 15\gev$.  After
optimization, 88\% of the FCNC signal events from the base selection
fall into the two signal regions, compared to 33\% of the background
events.  The inclusive signal acceptances for the decay $\ttbar\to
Wb\,Zc$ ($\ttbar\to Zc\,Zc$) are 0.43\% (0.58\%) for the $b$-tagged
selection, 0.34\% (0.86\%) for the non-$b$-tagged selection, and
0.10\% (0.16\%) for the control region.

%
%
To determine the FCNC branching fraction, we take into account single
or double FCNC decays of \ttbar pairs and normalize to the event yield
of a selection for the SM decay $\ttbar\to Wb\,Wb\to \ell \nu b\,
q\qbar' b$ (``lepton+jets'') requiring at least two jets to be
secondary vertex $b$-tagged~\cite{Sherman}.  In 1.9\invfb we observe
277~\ttbar candidate events, consistent with a production cross
section of $8.8\pm0.7\,(\mathrm{stat.})\unit{pb}$ assuming
$\btWb=100\%$.  If \tZq decays were present, these additional \ttbar
decays are less likely to be reconstructed in the lepton+jets mode,
resulting in a measured \ttbar production cross section smaller than
the actual cross section. We correct for this effect by modifying the
measured cross section based on the limit we set on \btZq.

%
%
We extract a limit on the branching fraction \btZq from a fit to the
mass \chisq distribution using templates constructed from the MC
simulated mass \chisq distributions of the FCNC signal and the SM
backgrounds (\Zj, SM \ttbar, and dibosons).  The normalization of the
dominant \Zj background is the most difficult to estimate from data
and MC simulations; therefore it is extracted from the fit. The SM
\ttbar background is normalized to the observed event yield in the
lepton+jets decay mode; backgrounds from diboson production are
normalized to their theoretical cross sections. Both contributions are
fixed in the fit.  The expected background from SM \ttbar production
and diboson production is $2.2 \pm 0.2$ ($3.2 \pm 0.2$) events for the
$b$-tagged (non-$b$-tagged) selection. The $b$-tagged, the
non-$b$-tagged, and the control region are fit simultaneously.  We
include systematic uncertainties due to the shapes of the signal and
background templates by allowing the templates to change shape via a
histogram interpolation technique (horizontal template
morphing)~\cite{Read:1999kh}. The uncertainty due to the experimental
jet energy scale (JES) is more than three times larger than any other
uncertainty, and the data are most consistent with a shift of the JES;
therefore we only include JES induced shape uncertainties in the
fitting procedure.

We use the signal-depleted control region to constrain the background
shape uncertainties without losing sensitivity to a small FCNC
signal. Additionally we use the number of \Zj events observed in the
control region, \zcont, to place a loose constraint on the number of
\Zj events in the two signal regions, \zsig. We constrain the ratio
$\Rsig=\zsig/\zcont$ to the value estimated by the MC simulation,
$\Rsig=0.51\pm 0.10$ at the nominal JES. The uncertainty is
conservatively estimated by varying the energy scales in the \alp MC
generator. We adjust \Rsig as a function of the JES shift \sigjes,
keeping the relative uncertainty of 20\%.  The absolute number of \Zj
background events remains unconstrained.  To reflect the constraint on
\Rsig in the template fit, we parameterize the number of \Zj events
passing the $b$-tagged and non-$b$-tagged signal selections, \ztag and
\zanti, as $\ztag=\ftag\cdot\Rsig\cdot\zcont$ and
$\zanti=(1-\ftag)\cdot\Rsig\cdot\zcont$, where \ftag is the fraction
of \Zj events passing the $b$-tagged signal selection.  In summary,
the unconstrained parameters of the template fit include the branching
fraction \btZq, \zcont, the tagging fraction \ftag, and the shift in
JES \sigjes.  We apply a Gaussian constraint on the ratio \Rsig. From
the template fit to the data we measure a branching fraction
$\btZq=-1.49\%$. The fit result is summarized in Table~\ref{table:fit}
and in Fig.~\ref{fig:moneyplot}.

\begin{table}[t]
\begin{center}
  \caption{\label{table:fit} Results of the fit to data.  From the
    ratio of the number of \Zj events in the signal and the control
    regions and the tagging fraction \ftag, we obtain
    $\ztag=13.5$~events and $\zanti=53.9$~events.}
\vspace{2mm}
\small

\begin{tabular}{l D{.}{.}{-1} c D{.}{.}{-1}} 
  \hline
  {\bf Fit Parameter} & \multicolumn{3}{c}{\bf Value} \\
  \hline
  Branching Fraction, \btZq (\%)                      &  -1.49 &$\pm$& 1.52 \\
  \hline
  $Z$+Jets Events in Control Region, \zcont        & 129.0  &$\pm$& 11.1  \\
  Ratio Signal/Control Region, \Rsig                &  0.52 &$\pm$&  0.07 \\ 
  Tagging Fraction, \ftag                          &   0.20  &$\pm$&  0.06  \\
  Jet Energy Scale Shift, $\sigma_\mathrm{JES}$        &  -0.74 &$\pm$&  0.43 \\ 
  \hline
\end{tabular}
\end{center}
\vspace*{-8mm}
\end{table}

We employ a Feldman-Cousins (FC) limit calculation framework that
includes the handling of systematic uncertainties~\cite{Rtop}. The FC
construction is based on applying the above template fits to simulated
experiments that are generated taking into account all known sources
of systematic rate and shape uncertainty and their correlations
(systematic uncertainties are discussed in the next paragraphs). We
expect a limit of ($5.0\pm2.2$)\% in the absence of signal.  From the
FC construction we find $\btZq < 3.7\%$ at 95\% C.L.

%
%
We have studied systematic uncertainties due to shape uncertainties of
the fitted templates, rate uncertainties (signal acceptance,
background rate), and due to the normalization to the lepton+jets
event yield. We treat the dominant source of template shape
uncertainties, the experimental JES, as a free fit parameter. To
account for further shape uncertainties, we measure the bias on the
fitted branching fraction \btZq in simulated experiments generated
assuming the second largest source of shape uncertainties, the MC
generator used for the \Zj background.  We simulate \Zj events for
which the vertex energy scale is varied by factors of two and assign
the measured relative branching fraction bias at $\btZq =3.7\%$ as a
systematic uncertainty (5.6\%).

The rate uncertainties, summarized in Table~\ref{table:systematics}
are divided into anti-correlated uncertainties, those which cause
migration of events between $b$-tagged and non-$b$-tagged selections,
and correlated uncertainties, those which simultaneously increase or
decrease both selections.  The signal acceptance systematic
uncertainties are evaluated for the ratio of the FCNC signal
acceptance to the acceptance for the event selection used in the
lepton+jets normalization mode. Background rate uncertainties affect
only the smaller SM \ttbar and diboson backgrounds. The rate of the
dominant \Zj background in the control region is a free parameter in
the template fit; therefore we do not assign systematic uncertainties
for this.

\begin{table}[t]
\begin{center}
  \caption{\label{table:systematics} Systematic uncertainties of the
    FCNC signal acceptance relative to the acceptance of the
    lepton+jets normalization mode and the background rate for the
    $b$-tagged and non-$b$-tagged selections.}
\vspace{2mm}
\small

\begin{tabular}{p{40mm}cD{.}{.}{-1}cc} 
  \hline
  {\bf Systematic} & \multicolumn{2}{c}{\bf Acceptance} & \multicolumn{2}{c}{\bf Background} \\
  {\bf Uncertainty} & {\bf b} & \multicolumn{1}{c}{\bf Non-b}  &{\bf b} & {\bf Non-b} \\
  \hline
  ID, Trig, and PDF              & 1.0 & 1.0 & 1.5 & 1.5 \\
  Initial/Final State Rad.       & 4.8 & 5.5 & --- & ---\\ 
  $Z$ Helicity                  & 3.4 & 3.6 & --- & ---\\ 
  \hline
  Total Correlated               & 6.2 & 6.1 & 6.2 & 6.2\\ 
  \hline
  $b$-Tagging                    & 5.6 & 16.1 & 3.2 & 2.5 \\ 
  \btZc vs. \btZu                & 4.5 &  4.5 & --- & --- \\ 
  \hline
  Total Anti-Correlated          & 7.2 & 16.7 & 3.2 & 2.5 \\ 
  \hline
\end{tabular}
\vspace*{-0.3in}
\end{center}
\end{table}

Normalization to the lepton+jets event yields in the \ttbar production
cross section analysis removes nearly all uncertainties depending
directly on luminosity. Many other systematic uncertainties also
partially cancel. We absorb the statistical and systematic
uncertainties of the \ttbar production cross section measurement as
part of the systematic uncertainty (7.8\%).

In conclusion, we have searched for the top quark flavor changing
neutral current decay \tZq in events with a $Z$ boson and four or more
jets using CDF Run II data corresponding to 1.9~$\mathrm{fb}^{-1}$ of
integrated luminosity. The data are consistent with the SM background
prediction, and we set an upper limit on the branching fraction \btZq
of 3.7\% at 95\% C.L. at a top quark mass of 175\gevcsq.  Assuming a
top quark mass of 170\gevcsq, the 95\% C.L.\ upper limit is 4.1\%.
Compared with the previous world's best limit of 13.7\% by the L3
experiment~\cite{L3fcnc} at LEP2, and the previous Tevatron upper
limit of 33\%, as reported by CDF Run~I~\cite{RunIfcnc}, our reported
limit of 3.7\% represents a substantial improvement.

%
%
We thank the Fermilab staff and the technical staffs of the
participating institutions for their vital contributions. This work
was supported by the U.S.\ Department of Energy and National Science
Foundation; the Italian Istituto Nazionale di Fisica Nucleare; the
Ministry of Education, Culture, Sports, Science and Technology of
Japan; the Natural Sciences and Engineering Research Council of
Canada; the National Science Council of the Republic of China; the
Swiss National Science Foundation; the A.P.\ Sloan Foundation; the
Bundesministerium f\"ur Bildung und Forschung, Germany; the Korean
Science and Engineering Foundation and the Korean Research Foundation;
the Science and Technology Facilities Council and the Royal Society,
UK; the Institut National de Physique Nucleaire et Physique des
Particules/CNRS; the Russian Foundation for Basic Research; the
Comisi\'on Interministerial de Ciencia y Tecnolog\'{\i}a, Spain; the
European Community's Human Potential Programme; the Slovak R\&D
Agency; and the Academy of Finland.

%
%

\def\Journal#1#2#3#4{{#1} {\bf #2}, #3 (#4)}
\def\NCA{Nuovo Cimento}
\def\NIM{Nucl.\ Instrum.\ Methods}
\def\NIMA{{Nucl.\ Instrum.\ Methods} A}
\def\NP{Nucl.\ Phys.} 
\def\NPB{{Nucl.\ Phys.} B}
\def\PLB{{Phys.\ Lett.}  B}
\def\PRL{Phys.\ Rev.\ Lett.}
\def\RPP{Rep.\ Prog.\ Phys.}
\def\PRD{{Phys.\ Rev.} D}
\def\PR{Phys.\ Rep.}
\def\PRP{Prog.\ Theor.\ Phys}
\def\ZPC{{Z.\ Phys.} C}
\def\MPL{{Mod.\ Phys.\ Lett.} A}
\def\EPJC{{Eur.\ Phys.\ J.} C}
\def\CPC{Comput.\ Phys.\ Commun.}

\renewcommand{\baselinestretch}{1}


\begin{thebibliography}{99}
\bibitem{cdfdefs} In the CDF geometry, $\theta$ is the polar angle
  with respect to the proton beam axis (positive $z$ direction), and
  $\phi$ is the azimuthal angle.  The pseudorapidity is $\eta=-\ln [
  \tan(\theta/2) ]$.
\end{thebibliography}
\end{document}